\documentclass[a4paper]{article}

\usepackage{INTERSPEECH2022}
\begin{document}
\maketitle

\def\thefootnote{*}\footnotetext{Equal contribution.}\def\thefootnote{\arabic{footnote}}
\begin{abstract}
Large-scale speech self-supervised learning (SSL) has emerged to the main field of speech processing, however, the problem of computational cost arising from its vast size makes a high entry barrier to academia.
In addition, existing distillation techniques of speech SSL models compress the model by reducing layers, which induces performance degradation in linguistic pattern recognition tasks such as phoneme recognition (PR).
In this paper, we propose FitHuBERT, which makes thinner in dimension throughout almost all model components and deeper in layer compared to prior speech SSL distillation works.
Moreover, we employ a time-reduction layer to speed up inference time and propose a method of hint-based distillation for less performance degradation.
Our method reduces the model to 23.8\% in size and 35.9\% in inference time compared to HuBERT.
Also, we achieve 12.1\% word error rate and 13.3\% phoneme error rate on the SUPERB benchmark which is superior than prior work.
\end{abstract} 

\raggedbottom
\section{Introduction}
Large-scale speech self-supervised learning (SSL) has emerged as an important field in speech processing recently due to its powerful performance and versatility.
Large amount of speech-only data can be utilized for pre-training, and even only small amount of paired data is adequate to fine-tune the model with great performance \cite{yi2020applying}.
HuBERT \cite{hsu2021hubert} and wav2vec 2.0 \cite{baevski2020wav2vec} both record word error rate (WER) of 1.8\%, which was the state-of-the-art performance on the test-clean LibriSpeech \cite{panayotov2015librispeech} benchmark. 
Not limited to the task of automatic speech recognition (ASR), speech SSL model can be expanded into various speech-related tasks by fine-tuning on a specific mainstream task such as automatic speaker verification (ASV) or keyword spotting (KS) {\cite{fan2020exploring, hussain2021multi}}. 

Despite its powerful performance and versatility, the main drawback of such wide and deep models is difficulty in usage due to its vast size.
The limitation of computational resource and time-consuming training caused by numerous parameters make speech SSL model usage burdensome.
According to the authors of \cite{baevski2020wav2vec}, they used total 128 V100 GPUs to pre-train wav2vec 2.0 \textsc{LARGE} for 2.3 days, which is mostly unavailable to academia.
These large-scale models require more memory and time at inference following high computational overhead during fine-tuning.

Knowledge distillation is a model compression technique which can be a possible solution for the above issue.
The knowledge from the cumbersome teacher model can be transferred to the student model by learning teacher's representation.
In \cite{sanh2019distilbert}, knowledge distillation is applied to BERT \cite{devlin2018bert}, one of the most prominent language representation model, by reducing the number of Transformer \cite{vaswani2017attention} layers and initializing with pre-trained BERT.
On the other hand, FitNets \cite{adriana2015fitnets} suggests thinner and deeper student than the teacher, matching not only the final outputs but also the intermediate representations as hints.
Authors in \cite{yim2017gift} make the student model imitate the relationship between features from two distinct layers by minimizing the corresponding distance of flow of solution procedure (FSP) matrices.

Meanwhile, in the field of speech SSL, few studies have been conducted using knowledge distillation techniques.
DistilHuBERT \cite{chang2021distilhubert} compresses 12 Transformer layers down to 2 by employing 3 distinct prediction heads.
In \cite{peng2021shrinking}, authors attempt to reduce the Transformer layers of wav2vec 2.0 \cite{baevski2020wav2vec} by introducing both KL-divergence and mean squared error (MSE) losses.
However, the main disadvantage of these two approaches is performance degradation of linguistic pattern recognition tasks such as ASR or phoneme recognition (PR). 

Accordingly, we propose a novel approach to student model design and distillation scheme, FitHuBERT, which can be applied to any Transformer-based speech SSL model.
In FitHuBERT, we design a model thinner and deeper compared to prior speech SSL distillation works.
A CNN feature extractor is designed in a channel-increasing manner with pointwise convolution.
For Transformer layers, dimensions of the self-attention and the inner-layer of feed-forward network (FFN) are reduced by 37.5\% and 84.3\%, respectively.
By using hint-based distillation and layer-wise prediction heads, FitHuBERT can be guided during the distillation for all Transformer layers.
Furthermore, a trainable time-reduction layer is introduced to attain faster inference.
We reduce the parameters of the model to 23.8\% and make the inference speed 2.8 times faster compared to the teacher model, HuBERT.
Also, performance of PR and ASR are relatively improved by 18.1\% and 9.6\% compared to DistilHuBERT respectively on the SUPERB benchmark \cite{yang21c_interspeech}.

While developing our method, a concurrent work \cite{wang2022lighthubert} with a similar goal proposes a two-stage distillation strategy, making use of pre-training distillation and large-sized Transformer supernet with neural architecture search.
Our approach, in contrast, explores a simple strategy of applying knowledge distillation directly to the pre-trained teacher model, having significantly lower training cost and simpler distillation strategy.

\begin{figure*}[!ht]
    \vspace{-1.4cm}
    \centering
    \resizebox{0.85\textwidth}{!}{\includegraphics{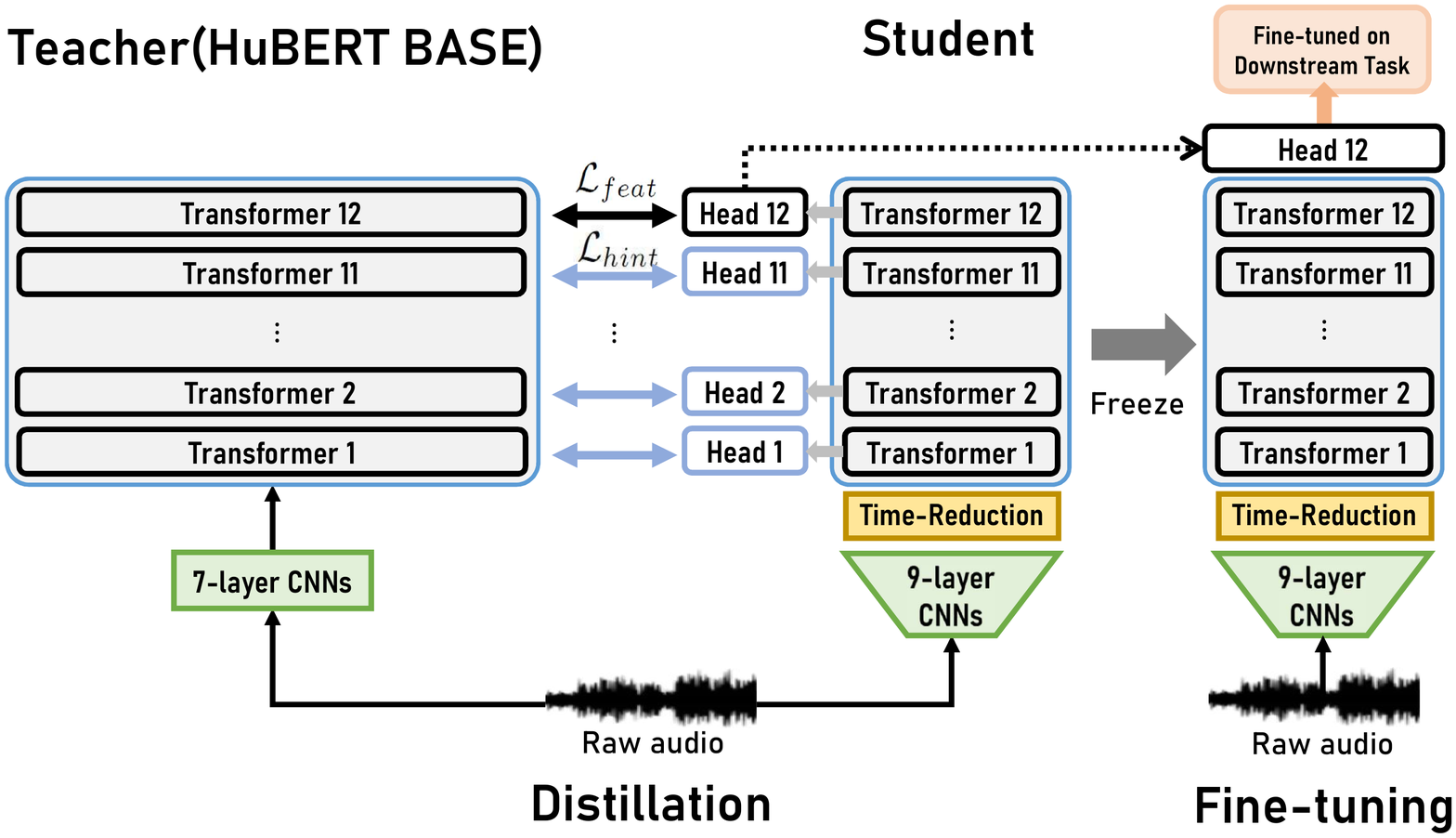}}
    \vspace{-1.25cm}
    \caption{The overall structure of FitHuBERT. It has a thin architecture in both CNNs and Transformers. The depth of Transformer layers is maintained as the same with HuBERT. Knowledge distillation is assisted with hints coming from the each layers of the teacher. Prediction heads solve the dimension mismatch between teacher and student. For fine-tuning, the prediction head at the only last layer is remained and utilized to downstream tasks.}
    \label{fig:FitHuBERT_diagram}
\end{figure*}

\section{Preliminaries}
\subsection{Speech SSL models}
In this paper, our proposed approach is based on HuBERT \cite{hsu2021hubert} and wav2vec 2.0 \cite{baevski2020wav2vec}, huge speech SSL models which achieved the state-of-the-art (SOTA) performance.
They are commonly composed of 7-layer CNN feature extractor with fixed number of channels, CNN-based positional embedding, and 12 consecutive Transformer layers for \textsc{BASE}, and 24 for \textsc{LARGE}.
The main difference between HuBERT and wav2vec 2.0 is in the pre-training stage where HuBERT separates the acoustic unit discovery step from the representation learning of masked prediction \cite{hsu2021hubert}. 
Evaluated on the SUPERB \cite{yang21c_interspeech} benchmark, HuBERT and wav2vec 2.0 topped all 10 benchmarks including ASR and PR.
Nonetheless, \textsc{BASE} models of HuBERT and wav2vec 2.0 have almost 100 millions of parameters, making them difficult to fine-tune and handle the inference time.
For these reasons, HuBERT and wav2vec 2.0 have been challenging to utilize in academia or individual researcher.

\subsection{SUPERB}
To verify the effectiveness of our distillation approach, we evaluate on the SUPERB benchmark \cite{yang21c_interspeech}, which consists of 10 different downstream tasks.
SUPERB is a benchmark for speech SSL model that aims wide use of speech representation in multiple speech-related tasks with minimal architecture changes and labeled data \cite{yang21c_interspeech}.
Pre-trained speech SSL model is fine-tuned on SUPERB downstream tasks by adding trainable lightweight layers on top while parameters of speech SSL model are frozen.
10 tasks included in SUPERB are PR, ASR, KS, ASV, query by example spoken term detection (QbE), speaker identification (SID), speaker diarization (SD), intent classification (IC), slot filling (SF) and emotion recognition (ER).

\begin{table*}[!]
\caption{Evaluation results on SUPERB. FBANK refers to log-mel filterbank. FitW2V2 is distilled wav2vec 2.0 in our approach. The metrics include number of parameters, inference time, PER (\%), WER (\%), maximum term weighted value (MTWV), F1 score (F1\%), concept error rate (CER\%), equal error rate (EER\%), and diarization error rate (DER\%). The best performance which outperformed DistilHuBERT are highlighted in bold. ASR performance is evaluated without language model.} 
\label{tab:main_results}
\resizebox{\textwidth}{!}{%
\begin{tabular}{@{}lcccccccccccc@{}}
\hline \hline
\toprule 
                          & $\#$ param & Inf. time & PR  & ASR  & KS  & QbE & SF  & IC           & ASV  & SD     & SID & ER  \\ \cmidrule(l){2-13}
Models                     & Millions$\downarrow$ & sec$\downarrow$ & PER$\downarrow$ & WER (w/o LM)$\downarrow$ & Acc$\uparrow$ & MTWV$\uparrow$ & F1$\uparrow$/ $ CER\downarrow$ & Acc$\uparrow$ & EER$\downarrow$ & DER$\downarrow$ & Acc$\uparrow$ & Acc$\uparrow$ \\ \midrule
\textbf{Baseline} \cite{yang21c_interspeech}
                          &          &                &     &     &     &     &     &               &      &        &     &     \\ \midrule
FBANK                      &     0    &     0     &82.01&23.18&8.63 &0.0058&69.64/52.94&9.1     &9.56  &10.55&8.5E-4&35.39     \\
TERA \cite{liu2021tera}    &  21.33   &     275.26     &49.17&18.17&89.48&0.0013&67.50/54.17&58.42   &15.89 &9.96 &51.51 &56.27 \\
HuBERT \textsc{BASE}       &    94.68 &     493.72     &5.41 &6.42 &96.30&0.0736&88.53/25.20&98.34   &5.11  &5.88 &81.42 &64.92     \\
wav2vec 2.0 \textsc{BASE}  &    95.04 &    490.49      &5.74 &6.43 &96.23&0.0233&88.90/24.77&92.35   &6.02  &6.08 &75.18 &63.43     \\ \midrule
\textbf{960h Distillation} &          &                &     &     &     &     &     &               &      &        &     &     \\ \midrule
DistilHuBERT \cite{chang2021distilhubert}
                           &    23.49 &     251.80     &16.27&13.37&95.98&0.0511&82.57/35.59&94.99   &8.55  &6.19 &73.54 &63.02    \\
\textbf{FitHuBERT-960 (ours)}              &  22.49   &     174.84    &13.32&12.09&\textbf{96.27}&0.0489&84.06/32.46&91.25   &8.00  &6.84 &55.71 &59.82    \\
\textbf{FitW2V2-960 (ours)}                &    31.63 &   176.82       &\textbf{12.22}&\textbf{11.44}&96.04&0.0475&\textbf{86.65/29.40}&93.38   &\textbf{6.65}  &6.44 &64.71 &62.35    \\ \midrule
\textbf{100h Distillation} &          &                &     &     &     &     &     &               &      &        &     &     \\ \midrule
\textbf{FitHuBERT-100 (ours)}              &    22.49 &   174.84      &14.05&12.66&96.23&\textbf{0.0579}&83.41/34.00&94.20   &7.88  &7.19 &54.24 &61.67    \\
\textbf{FitW2V2-100 (ours)}                &    22.49 &     174.84     &16.50&14.77&94.68&0.0380&81.95/34.74&90.03   &7.43  &6.94 &51.65 &62.87    \\ \bottomrule
\end{tabular}}
\end{table*}

\section{Methods}
In this section, we tackle the issues about the prior speech SSL distillation approaches, and demonstrate FitHuBERT.
The overall structure is depicted on Figure \ref{fig:FitHuBERT_diagram}.
\subsection{Model design}
\textbf{Maintaining Transformer layers} Main drawback of the prior works is the performance degradation of linguistic pattern recognition tasks such as ASR and PR compared to other tasks.
For example, relative PR performance drop of DistilHuBERT \cite{chang2021distilhubert} is about 200\% compares to the teacher.
Since the compression is done by reducing Transformer layers, depth of the student may not be enough to recognize various speech patterns.
This idea is supported by \cite{pham2019very} which claims thinner and deeper Transformer is better in ASR performance than wider yet shallower one.

To make thin Transformers, the dimensions of both self-attention and inner-layer of FFN in Transformer are reduced to have the same.
This aims to eliminate the bottleneck structure of FFN in Transformer, as argued in \cite{wu2020lite}.
Especially for short inputs, authors in \cite{wu2020lite} verifies that the computational cost of FFN occupies quite a large portion in Transformer.
With 50Hz downsampling and self-attention dimension of 768 in HuBERT \textsc{BASE} \cite{hsu2021hubert} and wav2vec 2.0 \textsc{BASE} \cite{baevski2020wav2vec}, about 15 seconds of speech also can be considered as a short input.
Therefore, assuming that the model processes short speech, there is no need to use too many parameters for maintaining the bottleneck structure.
%Accordingly, we reduce the dimensions of the self-attention and the inner-layer of FFN by 37.5\% and 84.3\%, respectively.

\noindent\textbf{Channel-increasing CNNs} Another problem is that the architecture of speech SSL does not consider the characteristics of CNN in speech field.
In speech, CNN is utilized to aggregate the speech features through time axis.
Many ASR studies exploit channel-increasing CNN for downsampling Mel-spectrogram, which implies that the large number of channels at the lower layers in the CNN is unnecessary 
\cite{hori2017advances, li2019jasper, kriman2020quartznet}.
This is analogous to the computer vision domain where the interchange in resolution and channel leads to more specialized CNN filters at the deeper parts in CNN \cite{zeiler2014visualizing}.

Considering this characteristics in the aspect of model compression, we suggest to reduce the number of channels in CNN at the lower layers.
In FitHuBERT, the channels of first CNN layer is reduced by factor of 4 compared to the teacher.
The channels get doubled whenever the kernel of convolution filter is decreased so that the final channels become equal to the teacher.
Furthermore, two pointwise convolutions are added before the channels get doubled.
% They are expected to facilitate sharing knowledge among channels through linear combination.

\subsection{Hint-based knowledge distillation}
Since FitHuBERT is smaller in dimension, it is more prone to unstable training or overfitting.
In addition, some studies have claimed that speech SSL models encode various acoustic and linguistic properties in specific Transformer layers \cite{pasad2021layer, chang2021exploration}.
Therefore, to transfer the knowledge from the teacher stably, distillation should be performed on every layer of the teacher, not only the last layer.
Distillations at the intermediate layers work as hints to guide the teacher's final representation, and provide more knowledge about the teacher's intermediate process \cite{adriana2015fitnets}.

The most intuitive way to implement this is to design the student with the same number of layers as the teacher, so that knowledge distillation can be done layer-to-layer.
For each Transformer layer, we attach a layer-wise prediction head to the output, where each head is composed of a temporal deconvolution layer \cite{zeiler2010deconvolutional} and a fully connected (FC) layer.
While training, these 12 distinct heads also have the role of matching the time length and self-attention dimension between the teacher and student. 
Time length mismatch is caused by the time-reduction layer which will be presented in the next subsection.
After distillation, prediction head on the last layer is only remained and used in the fine-tuning stage.

We use simple MSE loss to train the student by matching the representation of every layer in the teacher and student. 
The loss function for hint-based knowledge distillation is as follows:

\begin{equation}
    \mathcal{L}_{feat} = MSE(h^{(N)}_T, f_N(h^{(N)}_S)) ,
    \label{eqn:eq1}
\end{equation}

\begin{equation}
    \mathcal{L}_{hint} = \sum_{l = 1}^{N-1} MSE(h^{(l)}_T, f_l(h^{(l)}_S)) ,
    \label{eqn:eq2}
\end{equation}

\begin{equation}
    \mathcal{L}_{KD} = \mathcal{L}_{feat} + \lambda \mathcal{L}_{hint} ,
    \label{eqn:eq3}
\end{equation}
% 함수 및 벡터에 차원 간단하게 넣기
where $h_T$ and $h_S$ denote the Transformer layer representations of the teacher and student, respectively.
$N$ denotes the number of Transformer layers in the model.
$f$ represents the mapping function for the prediction head in each layer, which transforms the shape of representation.
$\lambda$ is selected by constant less than 1 in order to focus more on the last layer and avoid over constraints for distillation.  

\subsection{Time-reduction layer}
Maintaining the number of Transformer layers does not have a significant benefit in terms of inference time compared to the teacher model.
To tackle this issue, a trainable time-reduction layer consisting of simple temporal convolution is adopted \cite{haidar21_interspeech, chan2016listen} right before the Transformer layers.
Note that the self-attention module of each Transformer layer has $O(N^2)$ complexity where $N$ is the time length of the input \cite{vaswani2017attention}.
%The input sequence can be shrunk along time axis depending on the stride of the time-reduction layer.
The input sequence length can be shrunk to $N/k$ along the time axis depending on the time-reduction ratio $k$, the stride of the time-reduction layer.
Consequently, the computational complexity of self-attention reduces to a factor of $1/k^2$, making the inference of Transformer through all layers faster as well.

\section{Experiments}
\subsection{Implementation details}
\textsc{BASE} model of HuBERT \cite{hsu2021hubert} and wav2vec 2.0 \cite{baevski2020wav2vec} are the teacher model of FitHuBERT and FitW2V2 which have 7-layer CNNs with 12-layer Transformers in common.
Meanwhile, our student models have 9-layer CNNs with channel-increasing manner, and 12-layer Transformers with reduced dimensions of self-attention and inner-layer FFN.
For CNN layers, our models have channels of (128, 256, 256, 256, 256, 256, 512, 512, 512) with kernels of (10, 1, 3, 3, 3, 3, 1, 2, 2) and strides of (5, 1, 2, 2, 2, 2, 1, 2, 2).
For all Transformers, the dimensions of self-attention and inner-layer FFN is reduced to the same 480. 
Time-reduction ratio $k$ is set to 2.

Knowledge distillation is implemented with pytorch-lightning \cite{falcon2019pytorch} and fairseq \cite{ott2019fairseq}.
We exploit 100 hours of clean LibriSpeech dataset \cite{panayotov2015librispeech} for 100 epochs, and 960 hours of whole LibriSpeech for 80 epochs, which is represented as a suffix of -100 and -960, respectively.
A simple MSE loss is utilized for hint-based distillation with Adam optimizer and weight decay \cite{loshchilov2017decoupled}.
Optimizer is set to learning rate of $5\times10^{-4}$, warm-up proportion of 0.05, $\beta$ chosen from $[0.9, 0.98]$.
Epsilon and weight decay are all $1\times10^{-6}$.
FitHuBERT-960 and FitHuBERT-100 are trained with 2 NVIDIA 3080 Ti GPUs for 3.9 days and 0.5 days, respectively.
Batch size per GPU is 3 with 4 gradient accumulations.
$\lambda$ in Eq. (\ref{eqn:eq3}) is set to 0.1 to alleviate overly constrained distillation.

For fine-tuning, we follow the default fine-tuning settings of SUPERB \cite{yang21c_interspeech} benchmark for all 10 downstream tasks with a single NVIDIA 3080 Ti GPU.
Exceptionally, learning rate of SID task is set to $5\times10^{-3}$, and 2 GPUs are exploited for ASR.
Only when fine-tuning FitW2V2-960, all the prediction heads are remained and 12 trainable weights are additionally introduced.
The weighted sum of prediction head representations replaces the last layer prediction head.
Inference time is measured by averaging over 5 times on the test-clean of LibriSpeech dataset with a single batch.
Also, the performance of ASR is evaluated without using a language model (LM).

\subsection{Results}
Table \ref{tab:main_results} shows our main results, evaluated on the SUPERB benchmark.
%Our thinner and deeper student model design has better performance in linguistic pattern recognition tasks as intended.
Our student model design has better performance in linguistic pattern recognition tasks as intended.
The relative PR and ASR improvement of FitHuBERT-960 has improved by 18.1\% and 9.6\% respectively, compared to DistilHuBERT \cite{chang2021distilhubert}.
Surprisingly, FitHuBERT-100 performs better than DistilHuBERT on PR and ASR tasks which means FitHuBERT can learn the linguistic patterns faster with less data.
Performance of SF is also improved remarkably because the linguistic aspects of SF is similar with ASR, except for processing slot-types and slot-values \cite{yang21c_interspeech}.

Even if our model is designed with a purpose for particular downstream tasks, the utterance-level classification tasks such as SD or ER did not degrade much from the teacher model.
However, there was an obvious performance degradation for SID in both FitHuBERT and FitW2V2.
This shows that our models are slightly biased to linguistic pattern recognition tasks, focusing on recognizing local speech features rather than global features, which helps utterance-level classifications.
With this bias, the performance degradation is revealed when FitHuBERT or FitW2V2 tries to classify over 1000 different global features that underlies in a single utterance.
We emphasize that performance degradation is not severe for classifying global features with fewer classes such as ER or SD.

\section{Discussions}
In this section, we discuss the each component of our approach.
All distillations are performed using 100 hours of the train-clean LibriSpeech \cite{panayotov2015librispeech} dataset for 100 epochs.
After the distillation stage, models are fine-tuned and evaluated on the SUPERB benchmark with the tasks of ASR, IC, and SD.

\subsection{CNN architecture design}
To verify the effectiveness of our CNN structure, comparative experiments are conducted with various CNN architectures.
Table \ref{tab:CNN_comparison} shows that the pointwise convolution has contributed significantly to the performance improvements for channel-increasing CNNs.
Only using the design choice of channel-increasing CNNs is not enough to learn high dimensional information. However, the intermediate pointwise convolution allows to share the knowledge among the channels.
This allows the representation be expressed meaningfully on the high dimensional subspace.

\subsection{Number of layers for hints}
In Table \ref{tab:hint-based-table}, we evaluate the performance of our model using the different number of layers for hint-based knowledge distillation.
For every epoch, we randomly select 2 or 4 different Transformer layers from the teacher except for the last layer, and use only them for distillation.
Distillation with no hints is also conducted which is $\lambda = 0$ in \eqref{eqn:eq3}.
There is a clear tendency of performance improvement especially on IC and ASR, as the number of layers used for hint-based distillation are increased from 0 to 12.
Results suggest that deploying all hints for distillation is beneficial for providing more knowledge to student. 

\subsection{Trade-off of time-reduction layer}
Table \ref{tab:trfactor} shows a trade-off between inference time and overall performance according to the time-reduction ratio.
%Trade-off between the overall performance and $k$ can be seen from the results of using three different time-reduction ratios.
Note that the model without time-reduction layer where $k$ is set to 1 surpasses the original FitHuBERT-100 and even DistilHuBERT on ASR and IC tasks which shows the effectiveness of our model design and distillation scheme.
If the inference time is not a big issue rather than the number of parameters, one can simply exclude the time-reduction layer, to attain better performance.

\begin{table}[!bt]
\caption{Evaluation results of the different CNN architectures.}
\label{tab:CNN_comparison}
\resizebox{0.47\textwidth}{!}{%
\begin{tabular}{lrr|ccccc}
\hline
 & \multicolumn{1}{l}{} & \multicolumn{1}{l|}{}      & $\#$ param. & Inf. time & IC  & ASR & SD  \\ \cline{4-8} 
 & & \multicolumn{1}{c|}{CNN architecture} & Millions$\downarrow$ & sec$\downarrow$       & Acc$\uparrow$ & WER (w/o LM)$\downarrow$ & DER$\downarrow$ \\ \hline
\multicolumn{3}{l|}{\textbf{FitHuBERT-100}}                       &  22.49        &     174.84      & \textbf{94.20} & \textbf{12.66} & 7.19        \\
 & \multicolumn{2}{r|}{(-) pointwise conv} & \multicolumn{1}{c}{21.97} & \multicolumn{1}{c}{153.23} & \multicolumn{1}{c}{91.64} & \multicolumn{1}{c}{13.75} & \multicolumn{1}{c}{7.41} \\
 &                      & fixed ch 256               & 21.42         & 163.72         & 93.15          & 13.51          & 7.24   \\
 &              & fixed ch 512             & \multicolumn{1}{c}{24.69} & \multicolumn{1}{c}{237.39} & \multicolumn{1}{c}{91.99} & \multicolumn{1}{c}{13.68} & \multicolumn{1}{c}{\textbf{6.95}}
\end{tabular}%
}
\end{table}

\begin{table}[!bt]
\caption{Evaluation results of the different number of hints.}
\label{tab:hint-based-table}
\centering
\resizebox{0.4\textwidth}{!}{
\begin{tabular}{rccc}
\hline
                        & IC  & ASR & SD  \\ \cline{2-4} 
$\#$ of hints             & Acc$\uparrow$ & WER (w/o LM)$\downarrow$ & DER$\downarrow$ \\ \hline
No hints &  92.96  &  13.76  &  7.24  \\
% 2 layers & \multicolumn{1}{l}{} & \multicolumn{1}{l}{} & \multicolumn{1}{l}{} \\
2 hints &  93.41  &  13.48  &  7.23  \\
4 hints &  93.62  &  13.15  &  \textbf{6.94}  \\
\textbf{All hints} &  \textbf{94.20}  &  \textbf{12.66}  &  7.19
\end{tabular}%
}
\end{table}

\begin{table}[!bt]
\caption{Evaluation results of different time-reduction ratio ${k}$.}
\label{tab:trfactor}
\resizebox{0.47\textwidth}{!}{%
\begin{tabular}{lccccc}
\hline
    & $\#$ param. & Inf. time & IC  & ASR & SD  \\ \cline{2-6} 
${k}$                 & Millions$\downarrow$ & sec$\downarrow$       & Acc$\uparrow$ & WER (w/o LM)$\downarrow$ & DER$\downarrow$ \\ \hline
1         &      \textbf{21.11}    &     276.33      &  \textbf{95.12}   &  \textbf{12.00}   &   \textbf{7.10}  \\
2 & 22.49 & 174.84 & 94.20 & 12.66 & 7.19 \\
3          &     23.18     &       \textbf{157.81}    &   89.00  &  14.60   &    7.20
\end{tabular}%
}
\end{table}

\section{Conclusion}
We propose a novel approach to student model design and distillation scheme, FitHuBERT.
The dimensions of self-attention and inner-layer of FFN in Transformer layers are reduced by 37.5\% and 84.3\%, respectively.
Hint-based distillation is introduced to provide hints to guide the teacher's final representation during training.
Time-reduction layer is adopted for faster inference.
Evaluated on the SUPERB benchmark, FitHuBERT surpasses prior distillation scheme in linguistic pattern recognition task. 
Our approach reduces the teacher model to 23.8\% in size and is 2.8 times faster in inference time.
Also, we achieve 12.1\% WER and 13.3\% PER on the SUPERB benchmark which is superior than prior work.

\section{Acknowledgements}
This work was supported by the National Research Foundation of Korea (NRF) grant funded by the Korea government (MSIT) (No. 2021R1A2C1014044).
\bibliographystyle{IEEEtran}

\bibliography{template}

\begin{thebibliography}{10}
\providecommand{\url}[1]{#1}
\csname url@samestyle\endcsname
\providecommand{\newblock}{\relax}
\providecommand{\bibinfo}[2]{#2}
\providecommand{\BIBentrySTDinterwordspacing}{\spaceskip=0pt\relax}
\providecommand{\BIBentryALTinterwordstretchfactor}{4}
\providecommand{\BIBentryALTinterwordspacing}{\spaceskip=\fontdimen2\font plus
\BIBentryALTinterwordstretchfactor\fontdimen3\font minus
  \fontdimen4\font\relax}
\providecommand{\BIBforeignlanguage}[2]{{%
\expandafter\ifx\csname l@#1\endcsname\relax
\typeout{** WARNING: IEEEtran.bst: No hyphenation pattern has been}%
\typeout{** loaded for the language `#1'. Using the pattern for}%
\typeout{** the default language instead.}%
\else
\language=\csname l@#1\endcsname
\fi
#2}}
\providecommand{\BIBdecl}{\relax}
\BIBdecl

\bibitem{yi2020applying}
C.~Yi, J.~Wang, N.~Cheng, S.~Zhou, and B.~Xu, ``Applying wav2vec2. 0 to speech
  recognition in various low-resource languages,'' \emph{arXiv preprint
  arXiv:2012.12121}, 2020.

\bibitem{hsu2021hubert}
W.-N. Hsu, B.~Bolte, Y.-H.~H. Tsai, K.~Lakhotia, R.~Salakhutdinov, and
  A.~Mohamed, ``Hubert: Self-supervised speech representation learning by
  masked prediction of hidden units,'' \emph{IEEE/ACM Transactions on Audio,
  Speech, and Language Processing}, vol.~29, pp. 3451--3460, 2021.

\bibitem{baevski2020wav2vec}
A.~Baevski, Y.~Zhou, A.~Mohamed, and M.~Auli, ``wav2vec 2.0: A framework for
  self-supervised learning of speech representations,'' \emph{Advances in
  Neural Information Processing Systems}, vol.~33, pp. 12\,449--12\,460, 2020.

\bibitem{panayotov2015librispeech}
V.~Panayotov, G.~Chen, D.~Povey, and S.~Khudanpur, ``Librispeech: an asr corpus
  based on public domain audio books,'' in \emph{2015 IEEE international
  conference on acoustics, speech and signal processing (ICASSP)}.\hskip 1em
  plus 0.5em minus 0.4em\relax IEEE, 2015, pp. 5206--5210.

\bibitem{fan2020exploring}
Z.~Fan, M.~Li, S.~Zhou, and B.~Xu, ``Exploring wav2vec 2.0 on speaker
  verification and language identification,'' \emph{arXiv preprint
  arXiv:2012.06185}, 2020.

\bibitem{hussain2021multi}
S.~Hussain, V.~Nguyen, S.~Zhang, and E.~Visser, ``Multi-task voice activated
  framework using self-supervised learning,'' \emph{arXiv preprint
  arXiv:2110.01077}, 2021.

\bibitem{sanh2019distilbert}
V.~Sanh, L.~Debut, J.~Chaumond, and T.~Wolf, ``Distilbert, a distilled version
  of bert: smaller, faster, cheaper and lighter,'' \emph{arXiv preprint
  arXiv:1910.01108}, 2019.

\bibitem{devlin2018bert}
J.~Devlin, M.-W. Chang, K.~Lee, and K.~Toutanova, ``Bert: Pre-training of deep
  bidirectional transformers for language understanding,'' \emph{arXiv preprint
  arXiv:1810.04805}, 2018.

\bibitem{vaswani2017attention}
A.~Vaswani, N.~Shazeer, N.~Parmar, J.~Uszkoreit, L.~Jones, A.~N. Gomez,
  {\L}.~Kaiser, and I.~Polosukhin, ``Attention is all you need,''
  \emph{Advances in neural information processing systems}, vol.~30, 2017.

\bibitem{adriana2015fitnets}
R.~Adriana, B.~Nicolas, K.~S. Ebrahimi, C.~Antoine, G.~Carlo, and B.~Yoshua,
  ``Fitnets: Hints for thin deep nets,'' \emph{Proc. ICLR}, pp. 1--13, 2015.

\bibitem{yim2017gift}
J.~Yim, D.~Joo, J.~Bae, and J.~Kim, ``A gift from knowledge distillation: Fast
  optimization, network minimization and transfer learning,'' in
  \emph{Proceedings of the IEEE Conference on Computer Vision and Pattern
  Recognition}, 2017, pp. 4133--4141.

\bibitem{chang2021distilhubert}
H.-J. Chang, S.-w. Yang, and H.-y. Lee, ``Distilhubert: Speech representation
  learning by layer-wise distillation of hidden-unit bert,'' \emph{arXiv
  preprint arXiv:2110.01900}, 2021.

\bibitem{peng2021shrinking}
Z.~Peng, A.~Budhkar, I.~Tuil, J.~Levy, P.~Sobhani, R.~Cohen, and J.~Nassour,
  ``Shrinking bigfoot: Reducing wav2vec 2.0 footprint,'' \emph{arXiv preprint
  arXiv:2103.15760}, 2021.

\bibitem{yang21c_interspeech}
S.~wen Yang, P.-H. Chi, Y.-S. Chuang, C.-I.~J. Lai, K.~Lakhotia, Y.~Y. Lin,
  A.~T. Liu, J.~Shi, X.~Chang, G.-T. Lin, T.-H. Huang, W.-C. Tseng, K.~tik Lee,
  D.-R. Liu, Z.~Huang, S.~Dong, S.-W. Li, S.~Watanabe, A.~Mohamed, and
  H.~yi~Lee, ``{SUPERB: Speech Processing Universal PERformance Benchmark},''
  in \emph{Proc. Interspeech 2021}, 2021, pp. 1194--1198.

\bibitem{wang2022lighthubert}
R.~Wang, Q.~Bai, J.~Ao, L.~Zhou, Z.~Xiong, Z.~Wei, Y.~Zhang, T.~Ko, and H.~Li,
  ``Lighthubert: Lightweight and configurable speech representation learning
  with once-for-all hidden-unit bert,'' \emph{arXiv preprint arXiv:2203.15610},
  2022.

\bibitem{liu2021tera}
A.~T. Liu, S.-W. Li, and H.-y. Lee, ``Tera: Self-supervised learning of
  transformer encoder representation for speech,'' \emph{IEEE/ACM Transactions
  on Audio, Speech, and Language Processing}, vol.~29, pp. 2351--2366, 2021.

\bibitem{pham2019very}
N.-Q. Pham, T.-S. Nguyen, J.~Niehues, M.~M{\"u}ller, S.~St{\"u}ker, and
  A.~Waibel, ``Very deep self-attention networks for end-to-end speech
  recognition,'' \emph{arXiv preprint arXiv:1904.13377}, 2019.

\bibitem{wu2020lite}
Z.~Wu, Z.~Liu, J.~Lin, Y.~Lin, and S.~Han, ``Lite transformer with long-short
  range attention,'' \emph{arXiv preprint arXiv:2004.11886}, 2020.

\bibitem{hori2017advances}
T.~Hori, S.~Watanabe, Y.~Zhang, and W.~Chan, ``Advances in joint ctc-attention
  based end-to-end speech recognition with a deep cnn encoder and rnn-lm,''
  \emph{arXiv preprint arXiv:1706.02737}, 2017.

\bibitem{li2019jasper}
J.~Li, V.~Lavrukhin, B.~Ginsburg, R.~Leary, O.~Kuchaiev, J.~M. Cohen,
  H.~Nguyen, and R.~T. Gadde, ``Jasper: An end-to-end convolutional neural
  acoustic model,'' \emph{arXiv preprint arXiv:1904.03288}, 2019.

\bibitem{kriman2020quartznet}
S.~Kriman, S.~Beliaev, B.~Ginsburg, J.~Huang, O.~Kuchaiev, V.~Lavrukhin,
  R.~Leary, J.~Li, and Y.~Zhang, ``Quartznet: Deep automatic speech recognition
  with 1d time-channel separable convolutions,'' in \emph{ICASSP 2020-2020 IEEE
  International Conference on Acoustics, Speech and Signal Processing
  (ICASSP)}.\hskip 1em plus 0.5em minus 0.4em\relax IEEE, 2020, pp. 6124--6128.

\bibitem{zeiler2014visualizing}
M.~D. Zeiler and R.~Fergus, ``Visualizing and understanding convolutional
  networks,'' in \emph{European conference on computer vision}.\hskip 1em plus
  0.5em minus 0.4em\relax Springer, 2014, pp. 818--833.

\bibitem{pasad2021layer}
A.~Pasad, J.-C. Chou, and K.~Livescu, ``Layer-wise analysis of a
  self-supervised speech representation model,'' \emph{arXiv preprint
  arXiv:2107.04734}, 2021.

\bibitem{chang2021exploration}
X.~Chang, T.~Maekaku, P.~Guo, J.~Shi, Y.-J. Lu, A.~S. Subramanian, T.~Wang,
  S.-w. Yang, Y.~Tsao, H.-y. Lee \emph{et~al.}, ``An exploration of
  self-supervised pretrained representations for end-to-end speech
  recognition,'' \emph{arXiv preprint arXiv:2110.04590}, 2021.

\bibitem{zeiler2010deconvolutional}
M.~D. Zeiler, D.~Krishnan, G.~W. Taylor, and R.~Fergus, ``Deconvolutional
  networks,'' in \emph{2010 IEEE Computer Society Conference on computer vision
  and pattern recognition}.\hskip 1em plus 0.5em minus 0.4em\relax IEEE, 2010,
  pp. 2528--2535.

\bibitem{haidar21_interspeech}
M.~A. Haidar, C.~Xing, and M.~Rezagholizadeh, ``{Transformer-Based ASR
  Incorporating Time-Reduction Layer and Fine-Tuning with Self-Knowledge
  Distillation},'' in \emph{Proc. Interspeech 2021}, 2021, pp. 2102--2106.

\bibitem{chan2016listen}
W.~Chan, N.~Jaitly, Q.~Le, and O.~Vinyals, ``Listen, attend and spell: A neural
  network for large vocabulary conversational speech recognition,'' in
  \emph{2016 IEEE international conference on acoustics, speech and signal
  processing (ICASSP)}.\hskip 1em plus 0.5em minus 0.4em\relax IEEE, 2016, pp.
  4960--4964.

\bibitem{falcon2019pytorch}
W.~Falcon \emph{et~al.}, ``Pytorch lightning,'' \emph{GitHub. Note:
  https://github. com/PyTorchLightning/pytorch-lightning}, vol.~3, p.~6, 2019.

\bibitem{ott2019fairseq}
M.~Ott, S.~Edunov, A.~Baevski, A.~Fan, S.~Gross, N.~Ng, D.~Grangier, and
  M.~Auli, ``fairseq: A fast, extensible toolkit for sequence modeling,''
  \emph{arXiv preprint arXiv:1904.01038}, 2019.

\bibitem{loshchilov2017decoupled}
I.~Loshchilov and F.~Hutter, ``Decoupled weight decay regularization,''
  \emph{arXiv preprint arXiv:1711.05101}, 2017.

\end{thebibliography}

% \begin{thebibliography}{9}
% \bibitem[1]{Davis80-COP}\textbf{[}
%   S.\ B.\ Davis and P.\ Mermelstein,
%   ``Comparison of parametric representation for monosyllabic word recognition in continuously spoken sentences,''
%   \textit{IEEE Transactions on Acoustics, Speech and Signal Processing}, vol.~28, no.~4, pp.~357--366, 1980.
% \bibitem[2]{Rabiner89-ATO}
%   L.\ R.\ Rabiner,
%   ``A tutorial on hidden Markov models and selected applications in speech recognition,''
%   \textit{Proceedings of the IEEE}, vol.~77, no.~2, pp.~257-286, 1989.
% \bibitem[3]{Hastie09-TEO}
%   T.\ Hastie, R.\ Tibshirani, and J.\ Friedman,
%   \textit{The Elements of Statistical Learning -- Data Mining, Inference, and Prediction}.
%   New York: Springer, 2009.
% \bibitem[4]{YourName17-XXX}
%   F.\ Lastname1, F.\ Lastname2, and F.\ Lastname3,
%   ``Title of your INTERSPEECH 2022 publication,''
%   in \textit{Interspeech 2022 -- 23\textsuperscript{rd} Annual Conference of the International Speech Communication Association, September 18-22, Incheon, Korea, Proceedings, Proceedings}, 2022, pp.~100--104.
% \end{thebibliography}

\end{document}